\documentclass[showpacs,showkeys,preprintnumbers, amsmath,amssymb,preprint]{revtex4-1}
\usepackage{amssymb}
\usepackage{amsfonts}
\usepackage{amsmath}
\usepackage{graphicx}
\usepackage{dcolumn}
\usepackage{bm}
\usepackage{makeidx}
\usepackage[latin1]{inputenc}
\usepackage[english, english]{babel}
\usepackage[dvips]{hyperref}
\usepackage{pstricks}

\usepackage{graphicx}
\def\be{\begin{equation}}
 \def\ee{\end{equation}}
 \def\bea{\begin{eqnarray}}
 \def\eea{\end{eqnarray}}

\makeindex

\begin{document}
\title{Bounded orbits for photons as a consequence of extra dimensions}
\author{P. A. Gonz\'{a}lez}
\email{pablo.gonzalez@udp.cl}
\affiliation{Facultad de Ingenier\'{\i}a y Ciencias, Universidad Diego Portales, Avenida Ej\'{e}rcito Libertador 441, Casilla 298-V, Santiago, Chile.}
\author{ Marco Olivares }
\email{marco.olivaresr@mail.udp.cl}
\affiliation{ Facultad de Ingenier\'ia y Ciencias, Universidad Diego Portales,
Avenida Ej\'ercito Libertador 441, Casilla 298-V, Santiago, Chile.}
\author{Yerko V\'{a}squez.}
\email{yvasquez@userena.cl}
\affiliation{Departamento de F\'{\i}sica y Astronom\'{\i}a, Facultad de Ciencias, Universidad de La Serena,\\
Avenida Cisternas 1200, La Serena, Chile.}
\date{\today}

\begin{abstract}
In this work, we study the geodesic structure for a geometry described by a spherically symmetric
four-dimensional solution embedded in a five-dimensional space known as a brane-based spherically symmetric solution. Mainly, we have found that the extra dimension contributes to the existence of
bounded orbits for the photons, such as planetary and circular stable orbits that have not been observed for other geometries.
\end{abstract}

\maketitle

\tableofcontents

\newpage

\section{Introduction}


Extra-dimensional gravity theories have a long history that begins with an original idea propounded by Kaluza and Klein \cite{Kaluza:1921tu} as a way to unify the electromagnetic and gravitational fields, and nowadays finds a new realization within modern string theory \cite{Font:2005td, Chan:2000ms}. Among the higher-dimensional models of gravity, the five-dimensional Randall-Sundrum brane worlds models \cite{Randall:1999ee, Randall:1999vf} have garnered a great deal of attention in the last decade. In particular,  their second model \cite{Randall:1999vf}  can be described by a single brane, i.e., a 3+1-dimensional hyper-surface embedded  in a  higher-dimensional space-time, in which the matter and gauge fields  of the standard model are confined to it, while the gravitational field can propagate in the fifth dimension, which has an infinite size. From the cosmological point of view, brane worlds offer a novel approach to our understanding of the evolution of the universe, and they have exhibited interesting cosmological implications, as witnessed in the study of missing matter problems. They also provide a new mechanism to explain the acceleration of the universe based on modifications to general relativity (GR) instead of introducing an exotic content of matter  (for instance see Ref. \cite{Maartens:2003tw} and the references therein). On the other hand, the Dvali-Gabadadze-Porrati (DGP) brane world model \cite{Dvali:2000hr}, whose gravity behaves as four-dimensional on a short distance scale but shows a higher-dimensional nature at larger distances, has also attracted great interest in the last decade  \cite{delCampo:2007zj, Gregory:2007xy, Lue:2005ya}. This brane world model is characterized by the brane on which the fields of the standard model are confined, and contains the induced Einstein-Hilbert term. It also exhibits several cosmological features \cite{Dvali:2001gm, Deffayet:2001pu, Deffayet:2001uk, Saavedra:2009zz}. It has also been shown that the effective four-dimensional equations obtained by projecting the five-dimensional metric onto the brane acquire corrections to GR \cite{Maeda:2003ar}.  Additionally, brane worlds models with a non-singular or thick brane have been considered in the literature (see for instance \cite{Emparan:2000fn} and references therein). It is also worth mentioning that 
the classic GR tests have been examined
for various spherically symmetric static vacuum solutions of brane world
models, for instance see \cite{Youm:2001qc, Bohmer:2009yx, CuadrosMelgar:2009qb, Zhou:2011iq, Zhao:2012zzc}. However, it should also be noted that the geometric localization mechanism implies a  four-dimensional mass for the photon \cite{Alencar:2015rtc}. 
Besides, 
a Schwarzschild four-dimensional
solution
can be embedded in a five-dimensional space, which is known as a brane-based spherically symmetric solution, and is given by the following metric:
\begin{equation}\label{metric1}
 ds^2=-\left(1-{2m\over \tilde{r}}\right)(d\tilde{x}^{4})^2+
 \frac{d\tilde{r}^2}{1-{2m\over \tilde{r}}}+\tilde{r}^2d\tilde{\theta}^2+
 \tilde{r}^2\sin^2\tilde{\theta}\,d\tilde{\phi}^2
 +(d\tilde{x}^{5})^2,
\end{equation}
where $\tilde{x}^5$ stands for the extra dimension, and represents the simplest example of a black string. However,  it was shown by Gregory and Laflamme that it is unstable under linear metric perturbations \cite{Gregory:1993vy}.
Additionally, the line element is conformally related to the five-dimensional black cigar solution 
\cite{Chamblin:1999by}. Moreover,  by adding an extra flat dimension to the Kerr solution of GR and performing a boost in the fifth dimension to this five-dimensional metric and then compactifying the extra dimension, a new four-dimensional charged spherically symmetric black hole was obtained together with a Maxwell and dilaton field in Ref. \cite{Frolov}; the hidden symmetries and geodesics of Kerr space-time in Kaluza-Klein theory were studied in Ref. \cite{Aliev:2013jya}. 

In this work, we consider  the geometry described by (\ref{metric1}). Then, we study analytically the geodesic structure for particles and discuss the different kinds of orbits for 
photons focusing on the effects of the extra dimension. It is important for any geometry to characterize the geodesic motion of massive particles or photons because it is possible to know the different trajectories of particles, which makes it possible   
to fix the free parameters of the theory. Additionally, the study of null geodesics has been used to calculate the absorption cross section for massless scalar waves at the high frequency limit or the geometric optic limit, because at the high frequency limit the absorption cross section can be approximated by the geometrical cross section of the black hole photon sphere $\sigma\approx\sigma_{geo}=\pi b_{u}^2$, where $b_{u}$ is the impact parameter of the unstable circular orbit of photons for an unbounded orbit. Moreover, in \cite{Decanini:2011xi, Decanini:2011xw} this approximation was improved at the high frequency limit by $\sigma \approx \sigma_{geo}+\sigma_{osc}$, where $\sigma_{osc}$ is a correction involving the geometric characteristics of the null unstable geodesics lying on the photon sphere, such as the orbital period and the Lyapunov exponent. 
It should be mentioned that in Ref. \cite{Cuzinatto:2014dka}, the authors considered the same geometry and  set up the mass parameter to be far below the value necessary for a black hole solution, and investigated 
how the parameter related to the extra dimension brings subtle but important corrections to the classic tests performed in GR, such as the perihelion shift of the planet Mercury, the deflection of light by the Sun, and the gravitational redshift of atomic spectral lines. Moreover, this parameter can be constrained in order to agree with the observational results. As we will show, in this paper we adopt the same constraint on the mass parameter and on the parameter related to the extra dimension in order to estimate the radius of the stable circular photon orbit around a star like the Sun.

This paper is organized as follows:  in Sec. \ref{STL} we present the procedure to obtain the equations of motion for neutral particles in the brane-based spherically symmetric solutions. In  Sec.  III we  give the exact solution for the circular orbits and we describe the analytical solution for the orbits with angular momentum in terms of the $\wp$-Weierstrass elliptic function. Then, in  Sec. \ref{RT}, we study the radial trajectories. Finally, in  Sec. \ref{summ} we conclude with some comments and final remarks.

\section{Geodesics}
\label{STL}
First, we consider a light-cone type transformation in spherical coordinates,
embedded in a five-dimensional space  \cite{Cuzinatto:2014dka} given by
\begin{equation}\label{e2}
r=\tilde{r},\qquad \theta=\tilde{\theta},\qquad \phi=\tilde{\phi},
\qquad x^4={\tilde{x}^4+\tilde{x}^5\over \sqrt{2}},
\qquad x^5={\tilde{x}^4-\tilde{x}^5\over  \sqrt{2}}.
\end{equation}
Thus, Eq. (\ref{metric1}) becomes
\begin{equation}\label{metric}
 ds^2= \frac{dr^2}{1-{2m\over r}}+r^2d\theta^2+
 r^2\sin^2\theta\,d\phi^2
 +{m\over r}(dx^{4})^2+{m\over r}(dx^{5})^2+
 2\left({m\over r}-1\right)dx^{4}dx^{5}.
\end{equation}
Now, with the aim of studying the motion of neutral particles around the brane-based spherically symmetric  solution, we derive the geodesic equations. 
So, the Lagrangian that allows to describe the motion of a neutral particle in the background
(\ref{metric}) is
\begin{equation}\label{e4}
  2\mathcal{L}=
  \frac{\dot{r}^2}{1-\frac{2m}{r}}
  +r^2\dot{\theta}^2
  +r^2\sin^2\theta\,\dot{\phi}^2
 +{m\over r}(\dot{x}^4)^2 +{m\over r}(\dot{x}^5)^2
  +2\left({m\over r}-1\right)\dot{x}^4\dot{x}^5
  =-\mu^2,
\end{equation}
where $\dot{a}=da/d\tau$, $\tau$ is an affine parameter along the geodesic that
we choose as the proper time, and $\mu$ is the test mass of the particle ($\mu=1$ for massive particles and $\mu=0$ for photons). 

The equations of motion are obtained from
$ \dot{\Pi}_{q} - \frac{\partial \mathcal{L}}{\partial q} = 0$,
where $\Pi_{q} = \partial \mathcal{L}/\partial \dot{q}$
are the conjugate momenta to the coordinate $q$, which yields
\begin{equation}\label{e5}
\dot{\Pi}_{r} =-{m\over r^2}\left(1-{2m\over r}\right)^{-2}\dot{r}^{2}
+r(\dot{\theta}^{2}+\sin^{2}\theta \dot\phi^2)
-{m\over 2\,r^2}(\dot{x}^{4}+ \dot{x}^5)^2,
\end{equation}
\begin{equation}\label{e6}
\dot{\Pi}_{\theta} = r^2\sin\theta \cos\theta \,\dot\phi^2,
\quad \dot{\Pi}_{\phi}=0,
\quad \dot{\Pi}_{x^4}=0
\quad\textrm{and}\quad \dot{\Pi}_{x^5}=0.
\end{equation}
So, we can observe that the conjugate momenta associated with the coordinates $\phi$, $x^4$, and $x^5$ are conserved. Therefore,
\begin{equation}\label{e7}
\quad \Pi_{r}= {\dot{r}\over 1-{2m\over r}},\quad
\Pi_{\theta} = r^{2}\dot{\theta} , \quad \Pi_{\phi}
= r^{2}\sin^{2}\theta\, \dot{\phi}
\end{equation}
\begin{equation}\label{e8}
\Pi_{x^4} = {m\over r}\dot{x}^4
  +\left({m\over r}-1\right)\dot{x}^5 , \quad
\textrm{and}\quad \Pi_{x^5}= {m\over r}\dot{x}^5
  +\left({m\over r}-1\right)\dot{x}^4.
\end{equation}
Now, without lack of generality we consider that the motion is developed on the invariant plane
 $\theta  = \pi/2$ and $\dot\theta =0$. So the above equations can be written as
 \begin{equation}\label{e9}
r^{2}\dot{\phi}\equiv h,\qquad
{m\over r}\dot{x}^4
  +\left({m\over r}-1\right)\dot{x}^5\equiv c_1\qquad
  \textrm{and}\quad   
{m\over r}\dot{x}^5
  +\left({m\over r}-1\right)\dot{x}^4\equiv c_2
\end{equation}
where $h$ is the angular momentum of the particle, and $c_1$ and $c_2$ are dimensionless integration constants. Thus, by defining $c_1+c_2\equiv-k$ and $c_1-c_2\equiv b$, we obtain
 \begin{equation}\label{e10}
\dot{x}^4={1\over 2}\left({k\over 1-{2m\over r}}+b\right)\qquad
  \textrm{and}\quad
  \dot{x}^5=
  {1\over 2}\left({k\over 1-{2m\over r}}-b\right) .
\end{equation}
Note that the coordinate $\tilde{x}^5$ can be obtained by subtracting  $\dot{x}_4$ and $\dot{x}_5$, using Eq. (\ref{e10}) we get
\begin{equation}
\dot{\tilde{x}}^5=\frac{\dot{x}^{4}-\dot{x}^{5}}{\sqrt{2}}=b \implies \tilde{x}^5=b \tau.
\end{equation}
Thus, we observe that the coordinate $x^{5}$ increases linearly with the affine parameter $\tau$. Therefore, the particles can escape to the additional dimension. However,  we consider that the extra dimension experienced by the fields is small, in order to viabilize a universal extra dimension such as in Ref. \cite{Cuzinatto:2014dka}. 
Finally, by substituting Eqs. (\ref{e9}) and (\ref{e10}) in Eq. (\ref{e4}), it follows that
\begin{equation}\label{e11}
\left(\frac{dr}{d\tau}\right)^2={k^2\over 2}-
\left(1-{2m\over r}\right)\left(\mu^2+{b^2\over 2}+{h^2\over r^2}\right).
\end{equation}
From this equation we observe that $k^2/2$ represents the energy of the particle.
Now, defining the new constant $\tilde{\mu}^2\equiv \mu^2+b^2/2$, 
the  effective potential $V(r)$ can be written as 
\begin{equation}\label{e12}
V(r)= \left(1-{2m\over r}\right)\left(\tilde{\mu}^2+{h^2\over r^2}\right),
\end{equation}

On the other hand, if we consider Eq. (\ref{e11}), the  orbit in polar coordinates is given by
\begin{equation}\label{e13}
\left(\frac{dr}{d\phi}\right)^2 = {r^4\over h^2}\left[{k^2\over 2}-
\left(1-{2m\over r}\right)\left(\tilde{\mu}^2+{h^2\over r^2}\right)\right].
\end{equation}
This equation determines the geometry of the geodesic. Moreover, the orbital Binet  equation yields
\begin{equation}\label{e14}
 \frac{d^2u}{d\phi^2} +u= {m\over h^2}\tilde{\mu}^2+3mu^2,
\end{equation}
where we have used the Keplerian change of variable $r$ to $u=1/r$.

\begin{figure}[!h]
 \begin{center}
  \includegraphics[width=100mm]{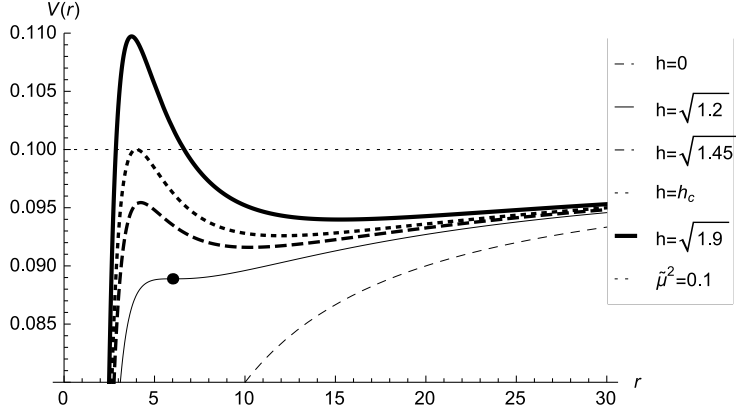}
 \end{center}
 \caption{Effective potential for photons with $m=1$, $\tilde{\mu}^2=0.1$, $h=0$ (thin-dashed line),  $h=\sqrt{1.2}$ (thin line), $h=\sqrt{1.45}$ (thick-dashed line), $h=h_c\approx 1.2649$ (thick-dotted line), and $h=\sqrt{1.9}$ (thick line).}
 \label{f1}
\end{figure}

Note that the effective potential $V(r)$ tends to $\tilde{\mu}^2$ when $r \rightarrow \infty$
  and that the constant $b$, associated with the extra dimension, also contributes to the effective potential. Also, observe that for photons $\mu=0$; however, $\tilde{\mu}^2=b^2/2$ is positive,  which implies that there are bounded orbits ($k^2/2<\tilde{\mu}^2$) for the photons as we  shall see in detail in the next section.  In the following, despite the geodesics being for photons and particles, we focus on the photon case ($\tilde{\mu}^2=0.1<1$). So,  in Fig. \ref{f1} we plot the effective potential $V(r)$ for photons with $m=0.1$, $\tilde{\mu}^2=0.1$ and for different values of the angular momentum of the particle $h$, where the point of inflexion corresponds to the last stable circular orbit (LSCO). Also, we can observe that there is a critical angular momentum of the particle $h_c\approx 1.2649$, where the energy of the unstable circular orbit takes the value $\tilde{\mu}^2$. For a $h<h_c$, the effective potential shows that all the unbound trajectories ($k^2/2\geq \tilde{\mu}^2$) can fall to the horizon or can escape to  infinity. If $h>h_c$, the maximum value of the effective potential is greater than $\tilde{\mu}^2$ and the particles with energy $\tilde{\mu}^2<k^2/2<k_c^2/2$ have return points.  The particles located on the right side of the potential barrier  that arrive from infinity have a point of minimum approximation and  are scattered to  infinity. However, particles located on the left side of the  potential barrier  have a return point from which they plunge to the horizon. On the other hand, the  minimum of the potentials corresponds to the stable circular orbits, whereas the  maximum corresponds to the unstable circular orbit.

\newpage

\section{Orbits}
Now, in order to obtain a full description of the motion of the neutral particles, we will find the geodesics analytically.
So, it is convenient to rewrite Eq. (\ref{e13}) as
\begin{equation}\label{tl11a}
  \left(\frac{dr}{d\phi}\right)^2=\frac{(\tilde{\mu}^2-k^2/2)}{h^2}\,r\,P(r),
\end{equation}
where the characteristic polynomial $P(r)$ is given by
\begin{equation}\label{tl12}
P(r)=-r^3+ \left({2m\tilde{\mu}^2\over \tilde{\mu}^2-k^2/2}\right)r^2-
\left({h^2\over \tilde{\mu}^2-k^2/2} \right)r+{ 2mh^2\over \tilde{\mu}^2-k^2/2}.
\end{equation}
Therefore, we can see that depending on the nature of its roots,
we can obtain the allowed motions for this configuration. Therefore, by integrating  Eq. (\ref{tl11a}),
we obtain the polar form of the orbit of the first kind for the neutral massive particles, which yields
\begin{equation}\label{c11}
r(\phi)={6m\over 24m\,\wp(\omega_0\mp\sqrt{2m}\,\phi; g_{2},g_{3})+1},
\end{equation}
where $\wp(x; g_2, g_3)$ is the $\wp$-Weierstrass elliptic function, with
the Weierstrass invariants given by
\begin{equation}\label{c12}
g_2 ={1 \over 48m^2}-{\tilde{\mu}^2 \over 4h^2},\qquad
g_3 ={1 \over 1728m^3}-{\tilde{\mu}^2 \over 96mh^2}
-{k^2-2\tilde{\mu}^2 \over 64mh^2},\qquad
\textrm{and}\quad
\omega_0=\wp^{-1}\left[{ 1\over 4R_0}-{ 1\over 24m}\right],
\end{equation}
with $\omega_0$ being an integration constant and $R_0$ the return point of the particle. Despite the test mass of the photon being null, in the following we define the  bounded and unbounded orbits if  $r$ remains bounded or not along the orbits, respectively. Also, the orbits of the first kind are defined as the relativistic analogues of the Keplerian orbits to which they tend in the Newtonian limit,  while the orbits of the second kind have no Newtonian analogues \cite{chandra}. 
\subsection{Bounded orbits}
 
\subsubsection{Circular orbits}
It is known that circular orbits ($r_{c.o.}$) correspond to  an extreme value of the potential, that is, $dV(r)/dr=0$. The last stable circular orbit occurs when the angular momentum is $h_{LSCO}=2\sqrt{3}\,m\,\tilde{\mu}$.
However, the minimum radius of a stable circular orbit is $r_{LSCO}=6\,m$.
Therefore, if $h<h_{LSCO}$, there are no circular orbits, and if $h>h_{LSCO}$ the circular orbits can be stable ($r_S$) or unstable ($r_U$), which yields
\begin{equation}\label{e15}
r_U={6m\,h^2 \over h_{LSCO}^2} 
\left(1-\sqrt{1-{ h_{LSCO}^2\over h^2 }}\right)\qquad
r_S={6m\,h^2 \over  h_{LSCO}^2} 
\left(1+\sqrt{1-{ h_{LSCO}^2\over h^2 }}\right).
\end{equation}
It  is important to mention that, despite the mass test of the photon being null, 
there are stable circular orbits for the light.  
Now, in order to estimate the radius of the circular orbits for photons, consider for example a circular orbit for photons around  a star like the Sun, without considering the effect of other bodies. The geometrical mass $m$ and the ratio $b/h$ are constrained to \cite{Cuzinatto:2014dka}:
\begin{equation}
m=1.4766250385 (1) km, \,\,\,\,\, \frac{b}{h} < 2.403015 \times 10^{-8} km^{-1}.
\end{equation}
Thus, considering $b/h \approx 2.403015 \times 10^{-8} km^{-1}$ from Eq. (\ref{e15}) we obtain the radius $r_{S} \approx 2.3 \times 10^{15} km$.

On the other hand, the periods for one complete revolution
of these circular orbits, measured in proper time and coordinate time ($x^4, x^5$), are
\begin{equation}\label{e16}
T_{\tau}= 2 \pi\sqrt{{{ r_{c.o.}^{3}-3mr_{c.o.}^{2}
} \over m\tilde{\mu}^2}},\qquad
T_{4}= 2 \pi\sqrt{ { r_{c.o.}^{3}
 \over 2m}}+{b\over 2}T_{\tau},
 \qquad
T_{5}= 2 \pi\sqrt{ { r_{c.o.}^{3}
 \over 2m}}-{b\over 2}T_{\tau}.
\end{equation}
Expanding the effective potential  around
to $r=r_S$, we can write
\begin{equation}\label{e17}
V(r)=V(r_S)+V'(r_S)(r-r_S)+{1\over2}V''(r_S)(r-r_S)^2+...,
\end{equation}
where $'$ means derivative with respect to the radial coordinate.
Obviously, in these orbits $V'(r_S)=0$. So, by defining the {\it smaller}
coordinate $x=r-r_S$, together with {\it the epicycle frequency}
$\kappa^2\equiv V''(r_S)/2$ \cite{RamosCaro:2011wx},  we can rewrite the above equation as
$V(x)\approx {k_S^2\over 2}+\kappa^2\,x^2$
where ${k_S^2\over 2}$ is the energy of the particle in the stable circular orbit.
Also, it is easy to see that test particles satisfy the harmonic equation of motion
$\ddot{x}=-\kappa^2\,x.$
Therefore, in our case, the epicycle frequency is given by
\begin{equation}\label{e20}
\kappa^{2}= {m\,\tilde{\mu}^2\over r_{S}^{3}}\left[{r_{S}-6m
 \over r_{S}-3m}\right].
\end{equation}
Note that this epicycle frequency is for photons, which  does not occur  in the Schwarzschild space-time.

\subsubsection{Orbits of the first kind, like planetary, and second kind} 
\begin{figure}[!h]
 \begin{center}
  \includegraphics[width=60mm]{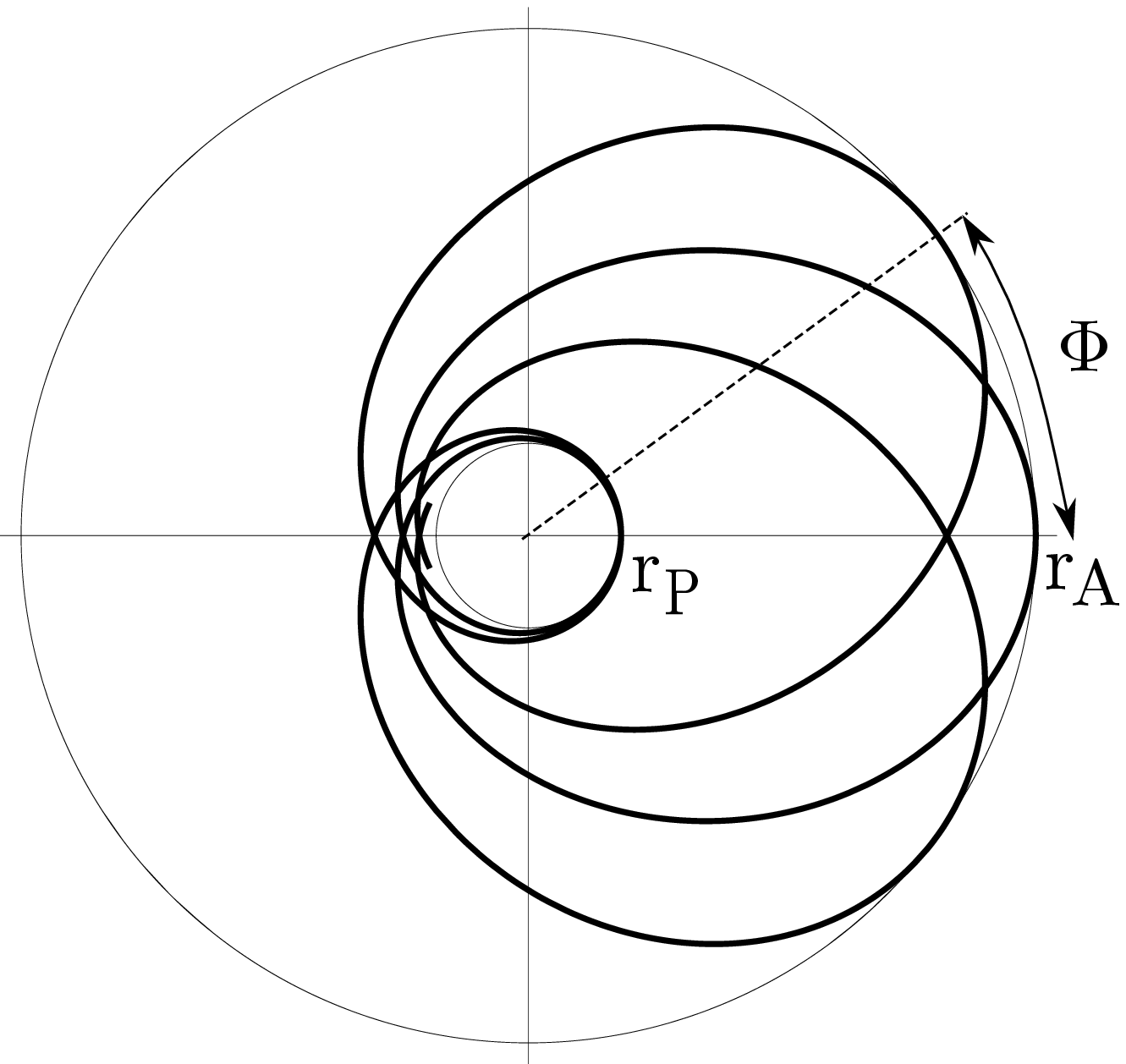}\,\,\,\,\,\,
   \includegraphics[width=60mm]{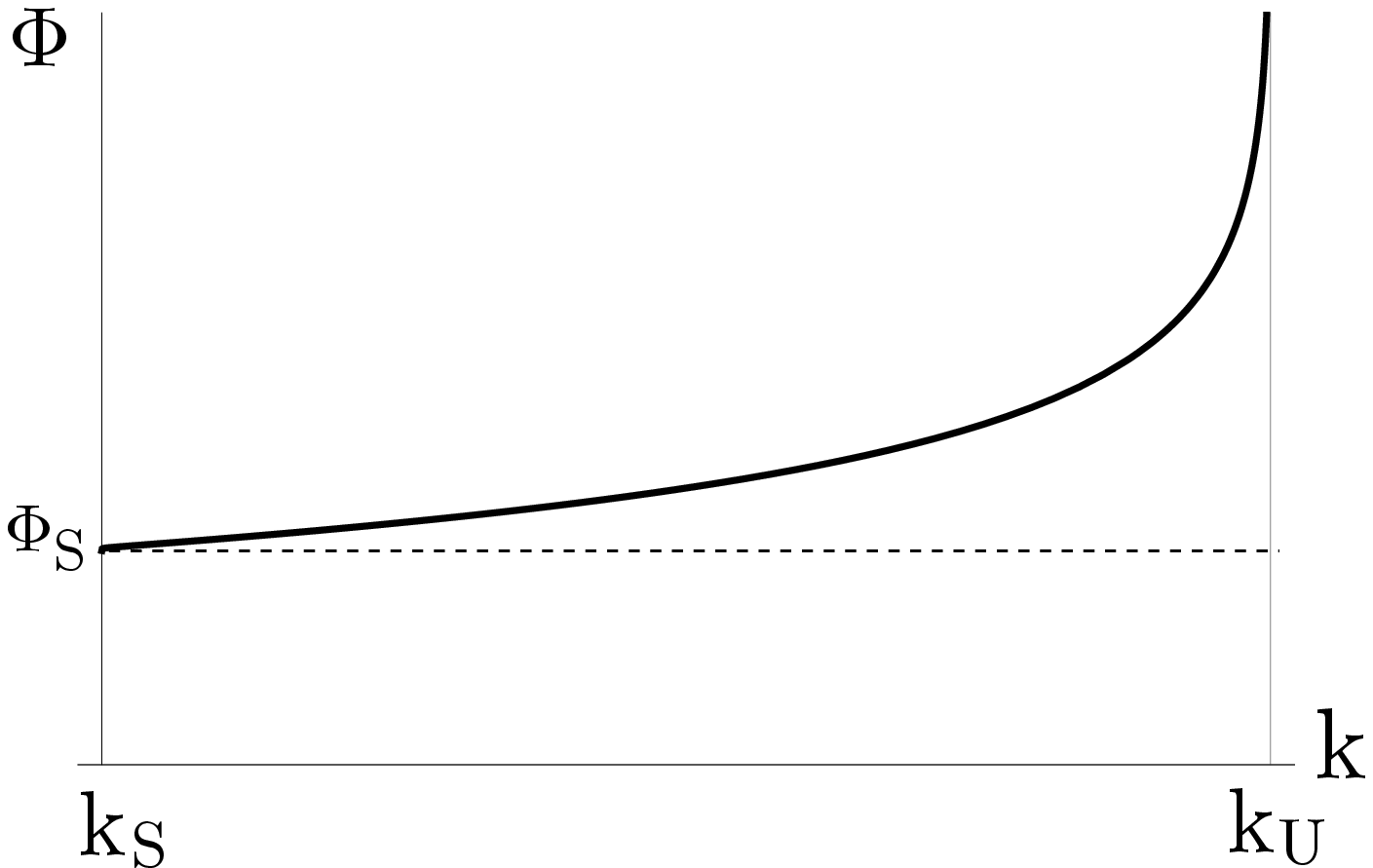}
 \end{center}
 \caption{The behavior of the bounded orbit with precession for photons is shown in the left figure for $m=1$, $h=1.20$, $k=0.435$, and $\tilde{\mu}^2=0.1$. The periastron corresponds to the small circle and it is located  at $r_P=7.09$ and the apoastron corresponds to the big circle and it is located at $r_A=16.21$. The variation of the precession angle  $\Phi$ relative to $k$ for $m=1$, $h=1.204$, and $\tilde{\mu}^2=0.1$ is shown in the right figure,  $k_S=0.42801$ and $\Phi_S=3.4374$ for the stable circular orbit, and $k_U=0.4369$ for the unstable circular orbit.}
 \label{f2}
\end{figure}

Orbits of the first kind occur when the energy lies  in the range $V(r_S)<k^2/2<V(r_U)<\tilde{\mu}^2$, 
and this case requires that $P(r)=0$ allows three real roots, all of which are positive;
and we write them as
\begin{equation}\label{c7}
r_{d}^{(\nu)}(k,h)={1 \over 3} 
\left({2m\tilde{\mu}^2\over \tilde{\mu}^2-k^2/2}\right)
+\sqrt{{\eta_2 \over 3}}
\cos \left[{1 \over 3} \arccos \left(3 \eta_{3}
\sqrt{{3 \over \eta_{2}^{3}}}\right)+
{2 \pi \nu \over 3}\right],\qquad (\nu=0, 1, 2)
\end{equation}
where
\begin{eqnarray}\label{c8}
  \eta_2  &=& 4\left[{1 \over 3}
  \left({2m\tilde{\mu}^2\over \tilde{\mu}^2-k^2/2}  \right)^2-
\left({h^2\over \tilde{\mu}^2-k^2/2} \right)\right], \\ \label{c9}
  \eta_3 &=& 4\left[{2 \over 27}
  \left({2m\tilde{\mu}^2\over \tilde{\mu}^2-k^2/2}\right)^3-
{1 \over 3}
\left({2m\tilde{\mu}^2\over \tilde{\mu}^2-k^2/2} \right)
\left({h^2\over \tilde{\mu}^2-k^2/2} \right)
+{ 2mh^2\over \tilde{\mu}^2-k^2/2}\right].
\end{eqnarray}
Thus, we can identify the apoastro distance as
$r_A=r_{d}^{(0)}$, and the periastro distance as $r_P=r_{d}^{(2)}$,
while the third solution can be recognized as the apoastro
distance to the orbits of the second kind, $r_F=r_{d}^{(1)}$. So, we can rewrite
the characteristic polynomial (\ref{tl12}) as
\begin{equation}\label{c10}
P(r)=(r_A-r)(r-r_P)(r-r_F).
\end{equation}
 Furthermore, we can determine the precession angle 
 corresponding to an oscillation, 
resulting in $\Phi=2\phi_P-2\pi$, where $\phi_P$ is the angle from the apoastro to the periastro. 
\begin{equation}\label{c13}
\Phi=\sqrt{{2\over m}}\left(
\wp^{-1}\left[{ 1\over 4r_P}-{ 1\over 24m}\right]
-
\wp^{-1}\left[{ 1\over 4r_A}-{ 1\over 24m}\right]
\right)-2\pi.
\end{equation}
Now, we can observe in Fig. \ref{f2} the behavior of the bounded orbit with precession, like planetary, for photons (left figure) 
and the variation of the precession angle $\Phi$ relative to $k$ 
(right figure),  where $k_S$ and $\Phi_S$ stands for the stable circular orbit, and $k_U$ for the unstable circular orbit. Note that the precession angle  increases when $k$ increases, in fact the precession angle tends to infinity when $k\rightarrow k_U$.

\begin{figure}[!h]
 \begin{center}
  \includegraphics[width=60mm]{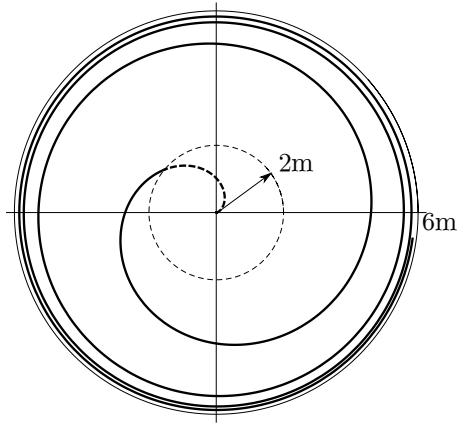}
 \end{center}
 \caption{The behavior of the last unstable critical orbit for photons. The small circle corresponds to the horizon, which is located  at $r=2m$. The apoastron distance is $r_{LSCO}=6m$ and corresponds to the big circle. The dashed line of the orbit mean that has not physical meaning inside the horizon. 
 Here, $m=1$, $h=1.109545$, $k=0.421637$, and $\tilde{\mu}^2=0.1$.}
 \label{f3}
\end{figure}

Also, if $h=h_{LSCO}$ the particle can orbit in  a stable circular orbit at $r_{LSCO} = 6m$. There is also a critical orbit  that approaches  the stable circular orbit asymptotically, see Fig. \ref{f3}. 

\begin{figure}[!h]
 \begin{center}
  \includegraphics[width=60mm]{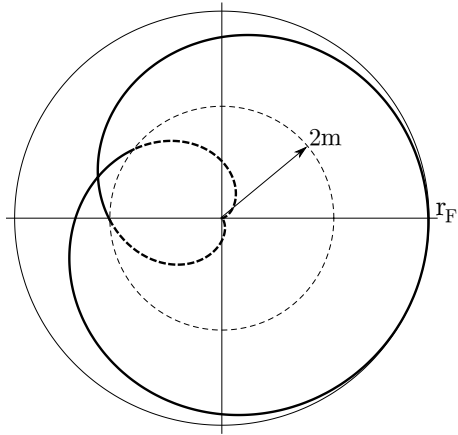}
 \end{center}
 \caption{Behavior of bounded orbit of the second kind for photons. The return point is located  at $r_F=3.77$, which corresponds to the big circle. The dashed small circle corresponds to the horizon, located  at $r=2m$. 
 Here $m=1$, $h=1$, $k=0.4$, and $\tilde{\mu}^2=0.1$.}
 \label{f4}
\end{figure}

In the second kind trajectory, the particle starts from a finite distance greater than the horizon and plunges towards the center (see Fig. \ref{f4}).


\newpage

\subsection{Unbounded orbits}

\subsubsection{Critical orbits}

\begin{figure}[!h]
 \begin{center}
  \includegraphics[width=70mm]{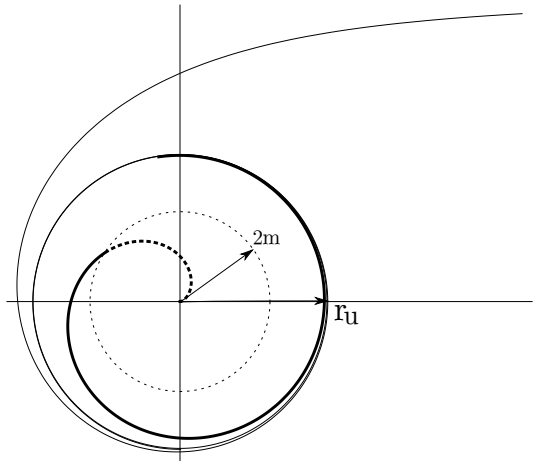}
 \end{center}
 \caption{Critical trajectories for photons with $m=1$, $\tilde{\mu}^2=0.1$, $h=2$, and $k_u=0.606$. For orbit of the first kind (thin line) the test particle arrived from infinity to $r_u=3.266$, where $r_u$ corresponds to the radius of the unstable circular orbit; and for orbits of the second kind (thick line) the test particle starts from $r_u$ and plunges into the horizon (dashed circle). 
  }
 \label{f7}
\end{figure}

The unstable circular orbits of radius $r=r_u$ corresponds to the maximum in the potential and is  allowed  for $h>h_c$ (see Fig. \ref{f1} for $h=\sqrt{1.9}$). In this case, the energy of the photon is $k_u^2/2=V(r_u)$. Also, there are two critical orbits that approach the unstable circular orbit asymptotically. In the first kind, the particle arises from infinity, and in the second kind, the particle starts from a finite distance greater than the horizon, but smaller than the unstable radius (see Fig. \ref{f7}).

\newpage

\subsubsection{Deflection of light}
\begin{figure}[!h]
 \begin{center}
  \includegraphics[width=60mm]{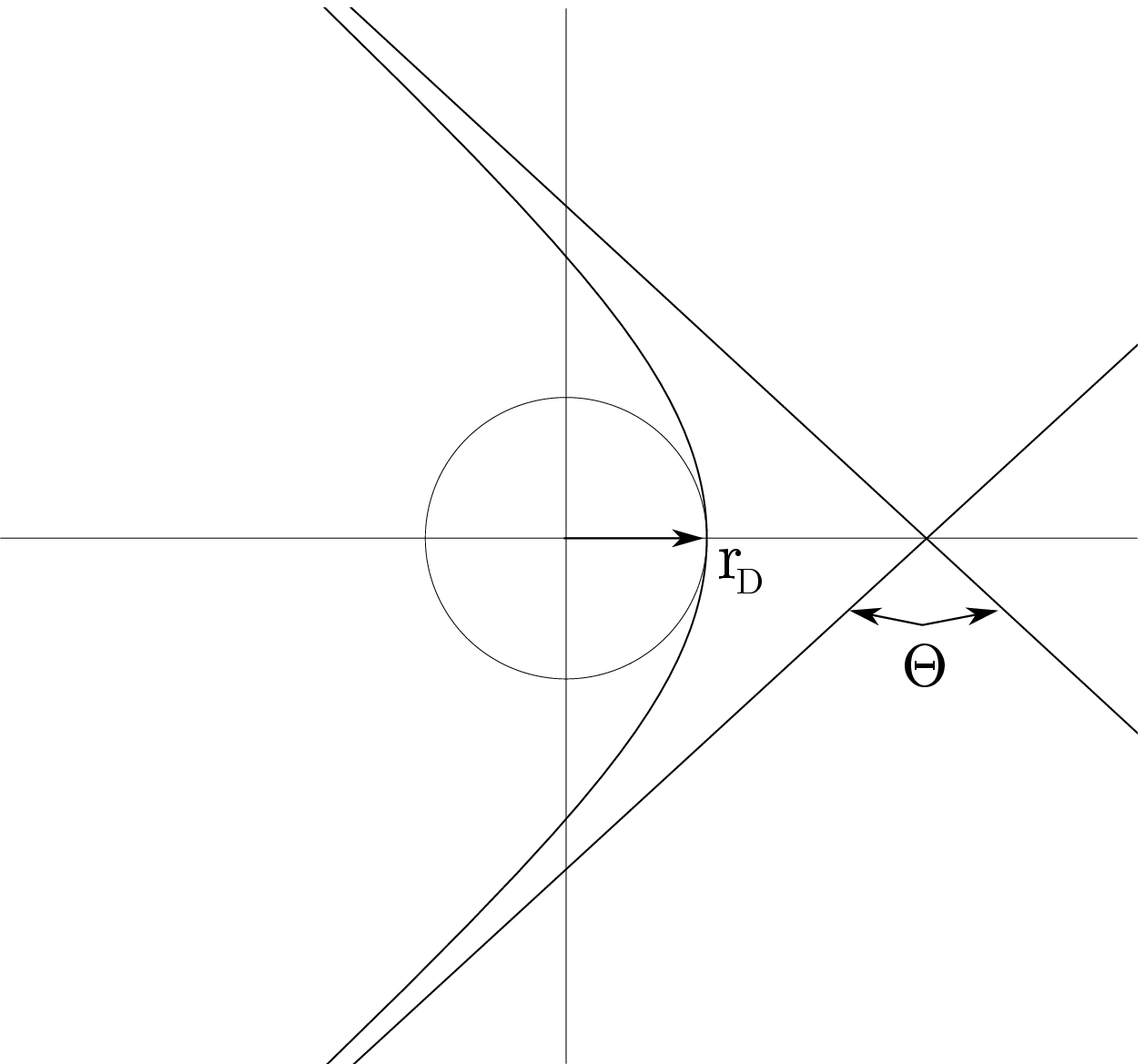}\,\,\,\,\,\,
   \includegraphics[width=60mm]{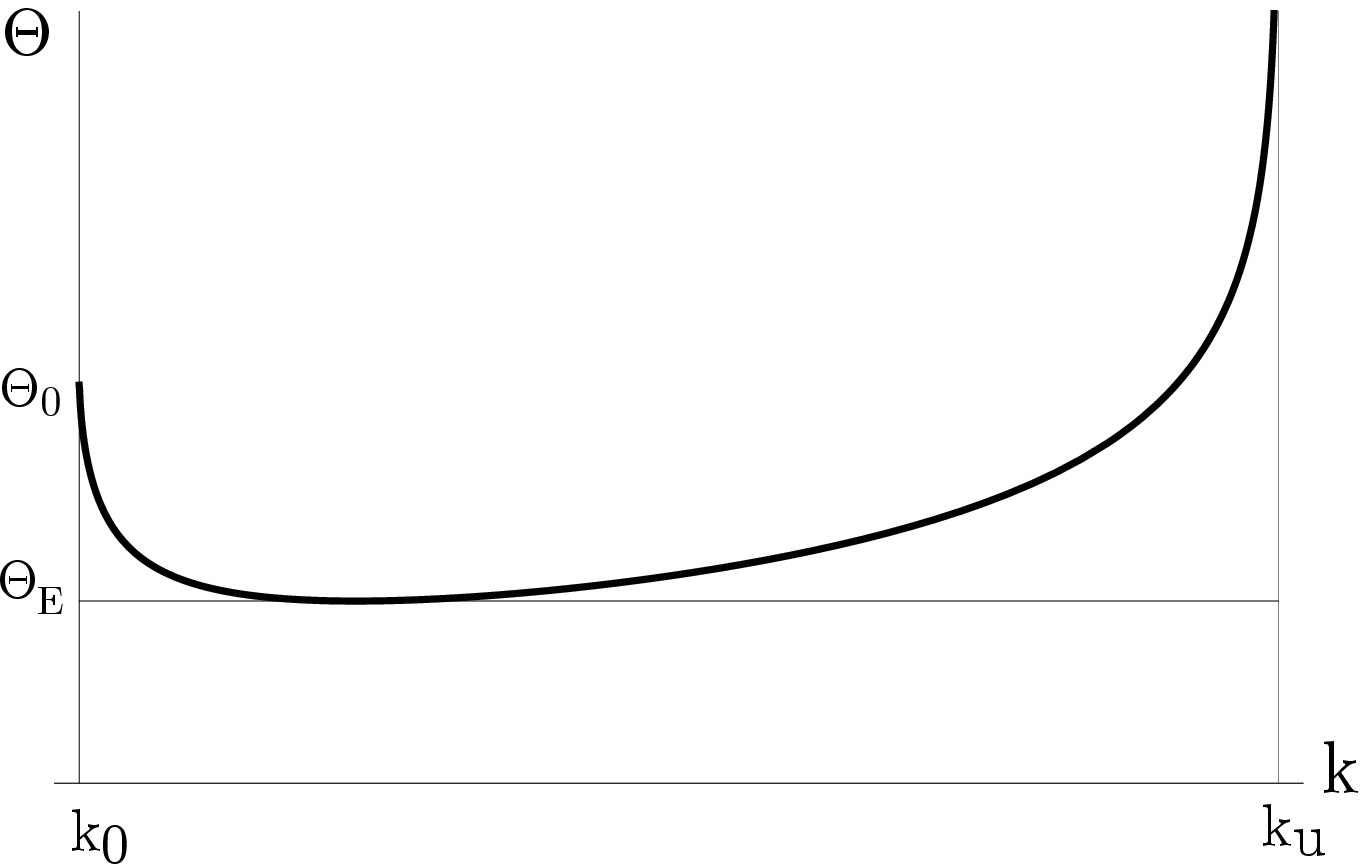}
 \end{center}
 \caption{The behavior of the deflection of light with $m=1$, $\tilde{\mu}^2=0.1$, $h=2$, $k=0.484$ is shown in  the left figure. $r_D=8.81$ corresponds to the closest distance and $\Theta$ corresponds to the deflection angle. The right figure shows the variation of the deflection angle $\Theta$ relative to $k$ for $m=1$, $\tilde{\mu}^2=0.1$, $h=2$. $\theta_0=3.72$ corresponds to the deflection angle for $k_0=\tilde{\mu}\sqrt{2}=0.447$, when $k^2/2=\tilde{\mu}^2$.  $\Theta_E= 1.65$ corresponds to the minimum deflection angle for $k_E=0.484$. For the unstable circular orbit $k_u=0.606$ and the deflection angle tends to infinity.}
 \label{f5}
\end{figure}

Orbits of the first kind occur when the energy lies in the range of $\tilde{\mu}^2<k^2/2<V(r_U)$,
and this case requires that $P(r)=0$ allows three real roots, which we can identify as  $r_D=r_d^{(0)}$, which correspond to the closest distance,  $r_f=r_d^{(2)}$ as an apoastro distance for the trajectories of the second kind  and the third  root, $r_3=r_d^{(1)}$ is negative without physical interest. Thus, we can rewrite
the characteristic polynomial (\ref{tl12}) as
\begin{equation}\label{c10}
P(r)=(r-r_D)(r-r_f)(r-r_3).
\end{equation}
Furthermore, we can determine the scattering angle, which is $\Theta=2\phi_{\infty}-\pi$
\begin{equation}\label{c13}
\Theta=\sqrt{{2\over m}}\left(
\wp^{-1}\left[{ 1\over 4r_D}-
{ 1\over 24m}\right]
-
\wp^{-1}\left[-{ 1\over 24m}\right]
\right)-\pi,
\end{equation}
where $\phi_\infty$ is the angle from the closest distance to infinity. In Fig. \ref{f5} we show the behavior of the deflection of light for the orbit with the closest distance (left figure), and 
we plot the variation of the deflection angle $\Theta$ relative to $k$ (right figure). 
Note that there is a minimum deflection angle $\Theta_E= 1.65$ for $k_E=0.484$ and 
$\Theta \rightarrow \infty$ when $k\rightarrow k_u$.

\subsubsection{Capture Zone}

\begin{figure}[!h]
 \begin{center}
  \includegraphics[width=80mm]{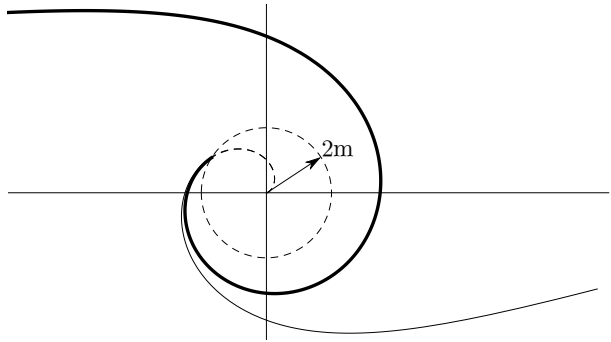}
 \end{center}
 \caption{Unbounded trajectories for  photons with $m=1$, $\tilde{\mu}^2=0.1$, $h=2$. For $k=0.61$ (thick line)  and $k=0.65$ (thin line), both trajectories for the photon arrive from infinity and plunge to the horizon (dashed circle). 
  }
 \label{f6}
\end{figure}

If $k>k_u$, the particle can escape to infinity or plunge into the horizon depending on the initial condition. The geometrical cross section of the black hole photon sphere $\sigma=\pi b_u^2$, where $b_u$ is the impact parameter of the circular orbit of photons, being $b_u=\frac{h}{k_u}\sqrt{2}$ (see Fig. \ref{f6}).

\newpage

\section{Radial trajectories} 
\label{RT}
The radial motion corresponds to
a trajectory with vanished angular momentum. Therefore, the effective potential 
for the particle is given by
\begin{equation} V(r)=\tilde{\mu}^2\left(1-{2m\over r}\right).
\label{i.13}\end{equation}
Observe that the potential for photons  does not  vanish as in the Schwarzschild case, where the potential for photons  is null. This implies that there are bounded radial trajectories, as we  shall see.

\subsection{Bounded trajectories:}
\begin{figure}[!h]
 \begin{center}
  \includegraphics[width=60mm]{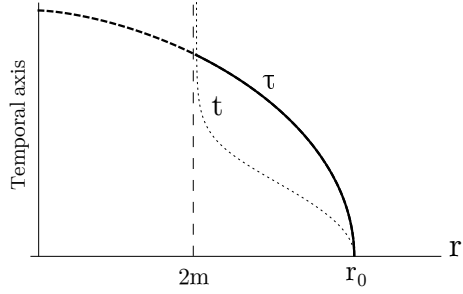}
 \end{center}
 \caption{The variation of the coordinate time ($\tilde{x}^4=t$) and the proper time $(\tau)$ along a bounded time-like radial geodesic described by a photon as test particle, starting at rest at $r_0=4$ and falling towards the singularity ($r=2m$), for $m=1$ and $\tilde{\mu}^2=0.1$. Continuous line for the proper time $\tau$ and dotted line for the temporal coordinate $\tilde{x}^4$. The dashed part of the curve inside the horizon has not physical meaning.}
 \label{f0121}
\end{figure}

It is possible to observe bounded trajectories if the condition $k^2/2<\tilde{\mu}^2$ is satisfied. The return point is given by
\begin{eqnarray}
r_0= {2m\tilde{\mu}^2\over \tilde{\mu}^2-k^2/2}.\label{w.6}\\
\end{eqnarray}
 The photons plunge into the horizon and the proper time $\tau(r)$ solution yields 
\begin{equation}
\tau(r)=\sqrt{{r_0\over 2m\tilde{\mu}^2}}
\left[r_0\,\arctan\sqrt{{r_0\over r}-1}
+r\,\sqrt{{r_0\over r}-1}\right],
\label{mr.1s}
\end{equation}
 and $x^4(r,b)$ reads
\begin{equation}
x^4(r,b)=\left({k\over2}+{b\over2}\right)\tau(r)
+{k\over2}\sqrt{{2mr_0\over\tilde{\mu}^2}}
\left[2\,\arctan\sqrt{{r_0\over r}-1}+\sqrt{{2m\over r_0-2m}}\ln\Omega(r)\right],
\end{equation}
where
\begin{equation}
\Omega(r)=\left|{\sqrt{r(r_0-2m)}+\sqrt{2m(r_0-r)}\over
\sqrt{r(r_0-2m)}-\sqrt{2m(r_0-r)}}\right|
\end{equation}
and $x^5(r,b)=x^4(r,-b)$. In Fig. \ref{f0121}, we plot the behavior of the proper time $\tau$ and the temporal coordinate $\tilde{x}^4=(x^4+x^5)/\sqrt{2}$, where the particle crosses the horizon in a finite proper time and the particle  takes an infinity coordinate time to reach the horizon. 

\subsection{Unbounded trajectories:   }
\begin{figure}[!h]
 \begin{center}
  \includegraphics[width=60mm]{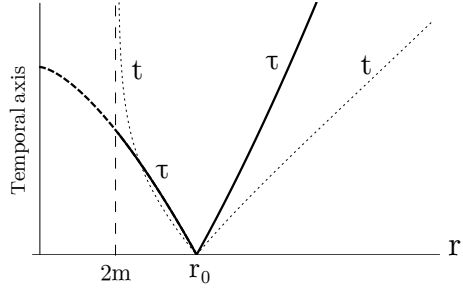}
 \end{center}
 \caption{The variation of the coordinate time ($\tilde{x}^4=t$) and the proper time $(\tau)$ along an unbounded time-like radial geodesic described by a photon as test particle, starting at rest at $r_0=4$ and falling towards the singularity ($r=2m$) or going towards infinity, for $m=1$ and $\tilde{\mu}^2=0.1$. Continuous line for the proper time $\tau$ and dotted line for the temporal coordinate $\tilde{x}^4$. The dashed part of the curve inside the horizon has not physical meaning.}
 \label{f0122}
\end{figure}

There are also unbounded trajectories if the condition $k^2/2\geq\tilde{\mu}^2$ is satisfied. In this case the proper time $\tau(r)$,  $x^4(r,b)$ and $x^5(r,b)$ yield
\begin{equation}
\tau(r)=\pm{2\over 3\sqrt{2m\tilde{\mu}^2}}
\left(r^{3/2}-r^{3/2}_0\right)
\end{equation}
\begin{align*}
\begin{split}
x^4(r,b)=\left({k\over2}+{b\over2}\right)\tau(r)
\pm{k\over2}\sqrt{{2m\over\tilde{\mu}^2}}
\left[2\,(\sqrt{r}-\sqrt{r_0})-\sqrt{2m}
\ln\left|{\sqrt{r}+\sqrt{2m}\over
\sqrt{r}-\sqrt{2m}}{\sqrt{r_0}-\sqrt{2m}\over
\sqrt{r_0}+\sqrt{2m}}\right|
\right]
\end{split} 
\end{align*}
$$
x^5(r,b)=x^4(r,-b),
$$
where, for the sake of simplicity, we have considered $k^2/2=\tilde{\mu}^2$.  In Fig. \ref{f0122}, we plot the behavior of the proper time $\tau$ and the temporal coordinate $\tilde{x}^4$, where the particle crosses the horizon in a finite proper time and the particle takes an infinity coordinate time to reach the horizon. 

 \newpage

\section{Summary}\label{summ}

In this manuscript, we have studied the geodesic structure for a geometry described by a spherically symmetric
four-dimensional solution embedded in a five-dimensional space known as a brane-based spherically symmetric solution, and we have described the different kinds of orbits for particles and photons.  Mainly, we have found that the effect of the extra dimension is to contribute to the effective potential through the parameter $b$.
This implies that there are bounded orbits for the photons, and we have found a stable circular orbit and the associated epicyclic frequency. Also, we have found bounded orbits that oscillate between an apoastro and a periastro distance, and we have determined the perihelion shift for the photons that have not been observed for other geometries. 
 In addition, we have found that the deflection of light is allowed for values of energy greater than $b^2/2$  and less than the energy for the unstable circular orbit. Note that this does not occur for the deflection of light  in a Schwarzschild space-time \cite{chandra}. The geometry analyzed  presents bounded and unbounded radial trajectories for neutral particles. However, it is possible to find bounded  trajectories for photons as a new behavior observed for null geodesics. In this sense, bounded orbits for the photons can be seen as a consequence of extra dimensions.


\begin{acknowledgments}
This work was partially funded by the Comisi\'{o}n
Nacional de Ciencias y Tecnolog\'{i}a through FONDECYT Grant 11140674 (PAG) and by the Direcci\'{o}n de Investigaci\'{o}n y Desarrollo de la Universidad de La Serena (Y.V.). P. A. G. acknowledges the hospitality of the Universidad de La Serena and Pontificia Universidad Cat\'{o}lica de Valpara\'{i}so, where part of this work was undertaken.
\end{acknowledgments}


\end{document}